# Schmidt decomposition and multivariate statistical analysis[1]

Yu. I. Bogdanov[2abc], N. A. Bogdanova[b], D. V. Fastovets[ab], V. F. Luckichev[a]

[a]Institute of Physics and Technology, Russian Academy of Sciences, Moscow, Russia
[b]National Research University of Electronic Technology MIET, Moscow, Russia
[c]National Research Nuclear University 'MEPHI', Moscow, Russia

## ABSTRACT

The new method of multivariate data analysis based on the complements of classical probability distribution to quantum state and Schmidt decomposition is presented. We considered Schmidt formalism application to problems of statistical correlation analysis. Correlation of photons in the beam splitter output channels, when input photons statistics is given by compound Poisson distribution is examined. The developed formalism allows us to analyze multidimensional systems and we have obtained analytical formulas for Schmidt decomposition of multivariate Gaussian states. It is shown that mathematical tools of quantum mechanics can significantly improve the classical statistical analysis. The presented formalism is the natural approach for the analysis of both classical and quantum multivariate systems and can be applied in various tasks associated with research of dependences.

## 1. INTRODUCTION

Schmidt decomposition [1] has widespread application in quantum information science [2]. This formalism is widely used in research of quantum optical systems, particularly, in analysis of spontaneous parametric down-conversion in a nonlinear crystal where pump photon breaks down into a photon pair [3-7]. Such correlated photon pairs are called biphotons. Schmidt correlations formalism allows us to analyze not only entanglement in quantum systems, but also classical correlations between subsystems in different probability distributions [8, 9].

In this article a special attention has been paid to the analytical form of Schmidt modes. One of the important results is development of an analytical theory for a quantum state, equivalent to the corresponding multivariate normal distribution.

A short description of Schmidt formalism is presented in section two. Definitions and properties of Schmidt decomposition, as well as the relation of this mathematical method to classical statistical analysis are also given in that section.

Calculation examples of Schmidt modes and Schmidt correlation coefficient for two-dimensional probability distributions are given in section three. In particular, the problem of photon beam separation by beam splitter has been considered. Correlation characteristics between two output modes according to input beam distribution have been studied. It has been shown that Schmidt correlation analysis is a more adequate method than the classical Pearson correlation analysis.

General approach to Schmidt correlation analysis in multivariate normal distribution is described in section four. Example of calculations with a four-dimensional Gaussian distribution is shown.

---

## 2. SCHMIDT DECOMPOSITION FORMALISM

The correlation data analysis method is based on Schmidt decomposition and complements the classical probability distribution study off a quantum state. Such formalism is the natural approach for the analysis of both classical and quantum correlations. The developed quantum analogue of the classical correlation analysis can be successfully applied to various tasks.

### 2.1 Schmidt decomposition

Let us have a two-particle system which is described by quantum state

$$|\psi\rangle = \sum_{j=1}^{s_A}\sum_{k=1}^{s_B} a_{jk} |j\rangle_A |k\rangle_B , \qquad (1)$$

where $s_A$ and $s_B$ - Hilbert space dimensions of subsystems $A$ and $B$ respectively. Then, there are orthonormal bases of subsystems spaces, for which the following Schmidt decomposition is valid:

$$|\psi\rangle = \sum_{k=1}^{r} \sqrt{\lambda_k} |\psi_k^A\rangle |\psi_k^B\rangle, \qquad (2)$$

where, $r \leq \min(s_A, s_B)$, $\lambda_k$ - nonnegative numbers, which are the Schmidt's coefficients (decomposition weights) satisfying normalization condition of total probability $\sum_k \lambda_k = 1$. $|\psi_k^A\rangle$ and $|\psi_k^B\rangle$ - Schmidt modes of corresponding subsystems. Schmidt coefficient allow us to estimate the effective number of modes in a decomposition by calculating the Schmidt number

$$K = \frac{1}{\sum_k \lambda_k^2}, \qquad (3)$$

This numerical characteristic describes the degree of nonrandomness of the matches. The higher the Schmidt number $K$ the less probable is the random match. At the minimum value $K = 1$ there is an a priori coincidence which marks absence of correlation. It turns out that Schmidt decomposition formalism – is an effective tool of correlation analysis.

### 2.2 Schmidt correlation coefficient

Schmidt correlation coefficient is introduced in [8] and is based on the Schmidt number (3)

$$\rho_{Schmidt}^2 = 1 - \frac{1}{K^2}, \qquad (4)$$

By complements a classical probability distribution to the quantum state [10], we can analyze the relationship between subsystems not only in the quantum state but using classical systems. A certain wave function may be assigned for an arbitrary probabilistic distribution $p(x_1, x_2, ..., x_n)$ which may be represented as follows:

$$\psi(x_1, x_2, ..., x_n) = \sqrt{p(x_1, x_2, ..., x_n)} \exp(iS(x_1, x_2, ..., x_n)), \qquad (5)$$

where $S(x_1, x_2, ..., x_n)$ - is the real-valued function acting as a phase. In the elementary case a zero function may be selected as the phase $S(x_1, x_2, ..., x_n) = 0$.

Let us consider two random correlated variables $X_1$ and $X_2$. We shall designate their mathematical expectations $m_1$ and $m_2$, dispersions $\sigma_1^2$ and $\sigma_2^2$ respectively, $\rho$ - is the correlation coefficient. Let the initial system be described by a probability distribution $P(x_1, x_2)$. We will specify that subsystem $A$ is formed by the random variable $X_1$, and subsystem $B$ - by the random variable $X_2$. If we consider that the random variables are normally distribute, then the Schmidt modes of subsystems $A$ and $B$ can be written in the following form [8]

$$\psi_k^A(x_1) = C_k^A H_k\left(\frac{x_1 - m_1}{\sigma_1}\sqrt{\frac{K}{2}}\right)\exp\left(-K\frac{(x_1-m_1)^2}{4\sigma_1^2}\right),$$

$$\psi_k^B(x_2) = C_k^B H_k\left(\frac{x_2 - m_2}{\sigma_2}\sqrt{\frac{K}{2}}\right)\exp\left(-K\frac{(x_2-m_2)^2}{4\sigma_2^2}\right),$$

(6)

where $C_k^A$ and $C_k^B$ - are normalization constants, $H_k(x)$ - are Hermite polynomials, and $K = \dfrac{1}{\sqrt{1-\rho^2}}$ - is the Schmidt number. In this case, the Schmidt weights $\lambda_k$ form a geometric progression with zero term equal to $\lambda_0 = \dfrac{2}{K+1}$, and the scale factor $q = \dfrac{K-1}{K+1}$.

To learn more about Schmidt decomposition and it relationship with correlation analysis please see article [9].

## 3. TWO-DIMENSIONAL ANALYSIS

A number of examples that follow show that Schmidt formalism reveals correlations which remain hidden by means of traditional analysis. Let us consider two problems associated with two-dimensional classical distributions analysis.

### 3.1 Research of noisy square dependence

Let us consider two random variables $X_1$ и $X_2$. This variables are interrelated by square dependence $X_2 = a(1-X_1^2)$, where $a$ - is a random value from range $0.9 \le a \le 1.1$. The borders of the region in Fig. 1 are described by the considered dependence with $a = 0.9$ and $a = 1.1$ respectively. We assume that a two-dimensional random variable $(X_1, X_2)$ has uniform distribution in the specified region.

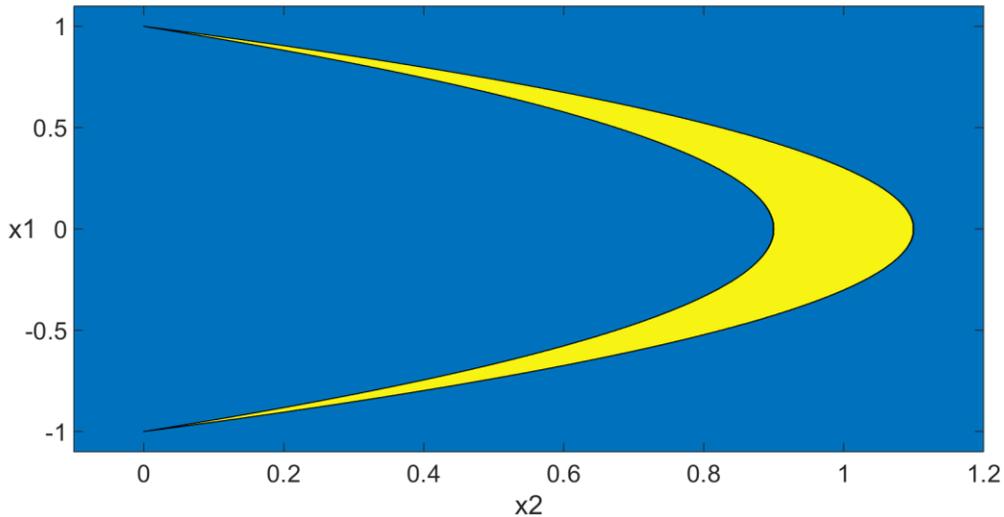

Figure 1. Evenly shaded area characterizes noisy square dependence $X_2 = a(1-X_1^2)$.

The Schmidt number for the distribution considered is approximately equal to $K = 4.4318$. The first three Schmidt modes are presented in Fig. 2. Their weights are respectively equal to $\lambda_0 = 0.4257$, $\lambda_1 = 0.1652$ and $\lambda_2 = 0.094$. The obtained values of Pearson and Schmidt correlation coefficients for the given distribution (Fig. 1) are as follows:

$$\rho^2_{Pearson} = 0,$$
$$\rho^2_{Schmidt} = 0.9491.$$

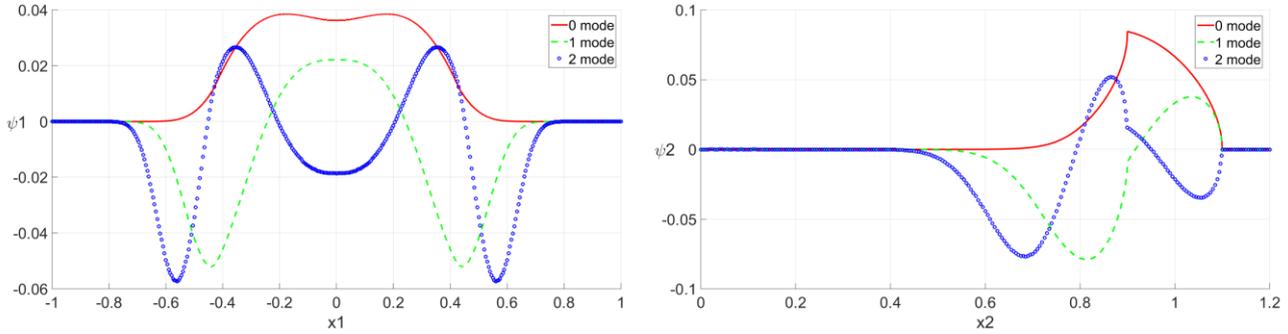

Figure 2. The first three Schmidt modes of the specified distribution (Fig. 1). To the left, the Schmidt modes corresponding to random variable $X_1$, and to the right, the Schmidt modes corresponding to random variable $X_2$.

Therefore, in this case, Schmidt decomposition makes it possible to reveal the correlation relationship which remains hidden in the traditional approach using Pearson correlation coefficient. This fact confirms the advantage of using of the Schmidt correlation coefficient in analysis of nongaussian distributions.

### 3.2 Photons correlation

The Schmidt correlation method can be applied to various tasks where it is necessary to explore degree of dependence between two systems. Let us consider the following problem associated with a real physical experiment: a beam of photons comes to the beam splitter (Fig. 3). The number of photons is given by a discrete random variable $k$ with the corresponding probability distribution $P(k)$.

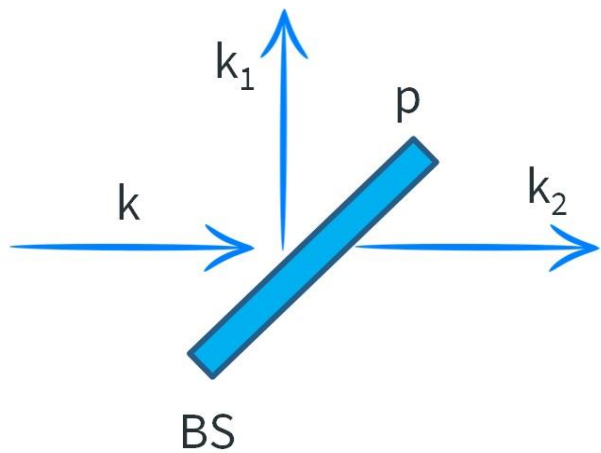

Figure 3. Flow diagram of the beam of photons through the beam splitter.

The beam splitter transmits a photon with specified probability $p$ and reflects the photon with probability $1-p$. At the output we have two beams with the numbers of photons $k_1$ and $k_2$ respectively. Therefore, the one-dimensional probability distribution $P(k)$ changes into a two-dimensional distribution $P(k_1,k_2)$. If the initial beam is described by the Poisson distribution, then the two-particle output distribution can be factorized on two Poisson distributions. Thus, there are no correlations between output channels. However, if we suppose that the parameter of Poisson distribution is a random variable which is described by gamma distribution, then the initial distribution will be given by a compound Poisson distribution

$$P(k) = \frac{a(a+1)...(a+k-1)\left(\frac{\mu}{a}\right)^k}{k!\left(1+\frac{\mu}{a}\right)^{a+k}}, \quad (7)$$

where, $a$ - is the clusterization parameter, $\mu$ - is the average number of photons. This distribution describes a multimode thermal state, where $a$ is the number of modes [11]. It can be shown that the same distribution also applies to the single-mode multiphoton-subtracted thermal state [12]. If the parameter of clusterization tends to infinity, then the compound Poisson distribution reduces to the common Poisson distribution. Now, the distribution of two output channels cannot be factorized on two distributions which describe two independent beams of photons. This amounts to an appearance of correlations in the system. Let us consider the initial beam of photons with the average number of photons equal to $\mu = 7$. Following the Schmidt formalism we shall calculate the correlation between the two beams of photons using different values of clusterization parameters and probabilities of transition through the beam splitter. These dependences are presented in Fig. 4.

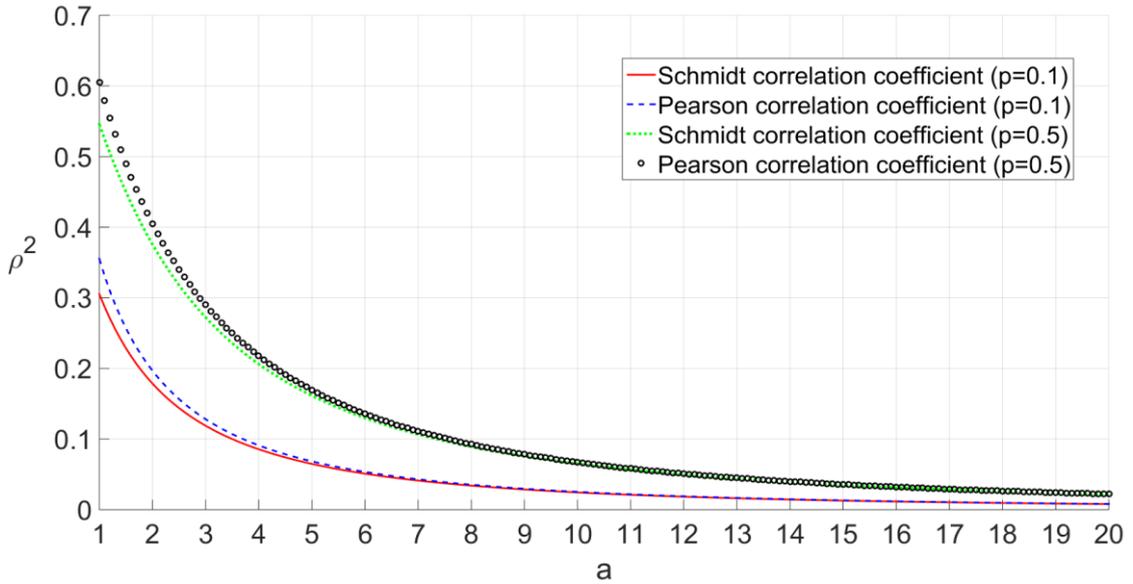

Figure 4. Dependence of the correlation coefficient on the parameter of clusterization with different transition probability values.

It can be shown that the Pearson correlation coefficient between the output beams of photons can be estimated by the following equation

$$\rho_{Pearson} = \frac{g^{(2)}-1}{\sqrt{\left(g^{(2)}-1+\frac{1}{p\mu}\right)\left(g^{(2)}-1+\frac{1}{(1-p)\mu}\right)}}. \quad (8)$$

Here, $g^{(2)}$ - is the correlation function of the second order. In our case: $g^{(2)} = 1 + \frac{1}{a}$.

As we can see in Fig. 4, the Schmidt correlation coefficient is quite different from the Pearson correlation coefficients for small values of clusterization parameter. As the average number of photons can take different values then we can study the behavior of the correlation coefficient for different values of parameters $a$ and $\mu$. These dependences are shown in Fig. 5. As we can see, the Pearson correlation coefficient differs from the Schmidt correlation coefficient in the area of small values of clusterization parameter.

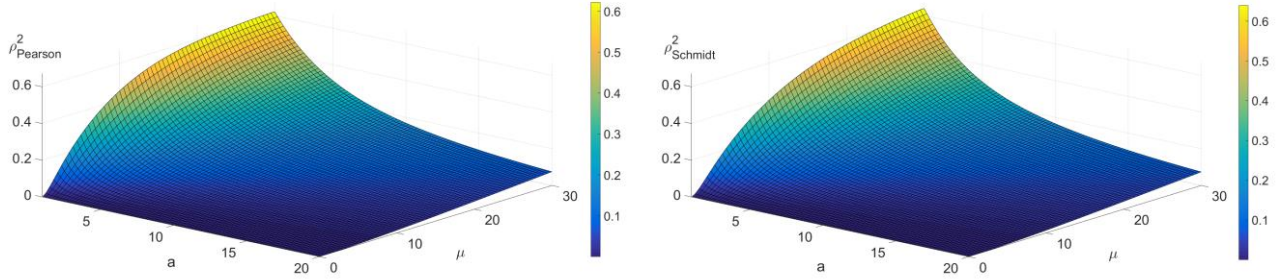

Figure 5. Dependences of Pearson (left) and Schmidt (right) correlation coefficient from parameter of clusterization and average number of photons. Transition probability throw beam splitter is equal to $p = 0.1$.

We have considered several examples from a wide range of possible applications of the Schmidt correlations formalism. The proposed approach can be successfully applied to the problems of image processing, classification of text data and analysis of different physical experiments.

## 4. SCHMIDT DECOMPOSITION FOR MULTIVARIATE GAUSSIAN STATE

### 4.1 Analytical equations of Schmidt modes

In this section we shall consider an important example – the case when Schmidt modes can be described analytically. It is the case of a multidimensional Gauss state.

Suppose we have a system with $p + q$ degrees of freedom, which is described by the vector

$$x = \left(x_1, ..., x_p, x_{p+1}, ..., x_{p+q}\right)^T, \tag{9}$$

A normal (Gaussian) state with zero mean for this system is given by:

$$\psi\left(x_1, ..., x_{p+q}\right) = \frac{1}{(2\pi)^{(p+q)/4} (\det \Sigma)^{1/4}} \exp\left(-\frac{1}{4} x^T \Sigma^{-1} x\right), \tag{10}$$

According to equation (5) we assume that the wave function is the root of the density of the corresponding normal distribution. Here $\Sigma$ - is the covariance matrix that has size $(p+q) \times (p+q)$. Let us consider that the first subsystem $A$ consists of the first $p$ variables, and the second subsystem $B$ consists of the other $q$ variables. To be specific let us assume that $p \leq q$.

It is necessary to calculate canonical coordinates and canonical correlation coefficients in order to find the Schmidt decomposition of Gaussian multivariate quantum state (10). The transformation to canonical coordinates $z$ and $w$ is given by

$$z = \sum_{j=1}^{p} \alpha_j x_j, \quad w = \sum_{j=1}^{q} \beta_j x_j. \tag{11}$$

The set of weights $\alpha$ and $\beta$ is selected so to maximize the correlation coefficient between the canonical variables. It is known that the canonical correlation coefficients are the roots of the following equation [13, 14].

$$\det\begin{pmatrix} -\lambda\Sigma_{11} & \Sigma_{12} \\ \Sigma_{21} & -\lambda\Sigma_{22} \end{pmatrix} = 0, \qquad (12)$$

where $\Sigma_{11}$ and $\Sigma_{22}$ - are symmetric covariance matrices, $\Sigma_{12}$ and $\Sigma_{21}$ - are cross-covariance matrices, $\Sigma_{12}^T = \Sigma_{21}$. It is assumed that the canonical variables have unit dispersions. As such, vectors $\alpha$ and $\beta$ satisfy a following normalization conditions:

$$\alpha^T \Sigma_{11} \alpha = 1, \quad \beta^T \Sigma_{22} \beta = 1. \qquad (13)$$

Let $A = \begin{pmatrix} 0 & \Sigma_{12} \\ \Sigma_{21} & 0 \end{pmatrix}$, $B = \begin{pmatrix} \Sigma_{11} & 0 \\ 0 & \Sigma_{22} \end{pmatrix}$. Let us consider a nontrivial solution for the following homogeneous equation

$$(A - \lambda B)b = 0. \qquad (14)$$

Note that the condition of existence of a nontrivial column vector $b$ is given by equation (12). Let us assume that covariance matrices $\Sigma_{11}$ и $\Sigma_{22}$ are nondegenerates. Let $c = B^{1/2} b$. Then this vector will satisfy the following condition $\left( AB^{-1/2} - \lambda B^{1/2} \right) c = 0$. Consequently

$$\left( B^{-1/2} A B^{-1/2} - \lambda I \right) c = 0. \qquad (15)$$

Where $I$ - is the identity matrix.

Therefore, the problem of finding the canonical correlation coefficient is reduced to calculating of eigenvalues of a symmetric matrix $A_1 = B^{-1/2} A B^{-1/2}$. All $p+q$ eigenvalues of $A_1$ are real, and $q-p$ values are equal to zero. All other values appear in pairs: $\pm\rho_1, ..., \pm\rho_p$. Let us choose strictly positive values $\rho_j$, where $j = 1, 2, ..., r$, and $r \leq p$ as the canonical correlation coefficients. Here $r$ - is the number of non-zero canonical correlation pairs. Each canonical correlation coefficient is associated with a Schmidt number

$$K_j = \frac{1}{\sqrt{1-\rho_j^2}}. \qquad (16)$$

The total Schmidt number is given by multiplication of all canonical partial Schmidt numbers.

$$K = \prod_{j=1}^{r} K_j, \qquad (17)$$

The obtained Schmidt number $K$ determines multiple correlation coefficients (4) between subsystems $A$ and $B$

$$\rho_{A \bullet B | Schmidt}^2 = 1 - \frac{1}{K^2}. \qquad (18)$$

Note that if we consider a normal distribution, than multiple correlation coefficients can also be calculated by the following standard equation [14, 15]

$$\rho_{A \bullet B}^2 = 1 - \frac{\det(\Sigma)}{\det(\Sigma_{11})\det(\Sigma_{22})}, \qquad (19)$$

where $\Sigma$ - is the initial covariance matrix. It is important to note that each canonical correlation coefficient $\rho_j$ generates is own set of Schmidt weights. This weights form a geometric progression. The parameters of this progression (the zero term and the scale factor) are given by equations: $\lambda_{0j} = \dfrac{2}{K_j + 1}$, $q_j = \dfrac{K_j - 1}{K_j + 1}$, $j = 1, 2, ..., r$.

The sets of coefficients $\alpha$ and $\beta$ for transform to the canonical variables are given by

$$\begin{pmatrix} -\rho_j \Sigma_{11} & \Sigma_{12} \\ \Sigma_{21} & -\rho_j \Sigma_{22} \end{pmatrix} \begin{pmatrix} \alpha \\ \beta \end{pmatrix} = 0, \; j = 1, 2, ..., r, \qquad (20)$$

or, what is equivalent, $\begin{pmatrix} \alpha \\ \beta \end{pmatrix}$ - eigenvector of matrix $\begin{pmatrix} -\rho_j \Sigma_{11} & \Sigma_{12} \\ \Sigma_{21} & -\rho_j \Sigma_{22} \end{pmatrix}$ with zero eigenvalue. The canonical correlation method allows us to generalize analytical equations of Schmidt modes (6) for arbitrary dimension systems. Note that canonical variables play the same role, as the normal modes in the problem of oscillation of coupled oscillators.

Consider the Hermitian modes of order $n_j$ ($\psi_{n_j}^A(z)$ and $\psi_{n_j}^B(w)$ respectively) for canonical pair number $j$, $n_j = 0, 1, ...$. Functions $\psi_{n_j}^A(z)$ and $\psi_{n_j}^B(w)$ can be calculated by the equations (6) with the Schmidt number $K_j$.

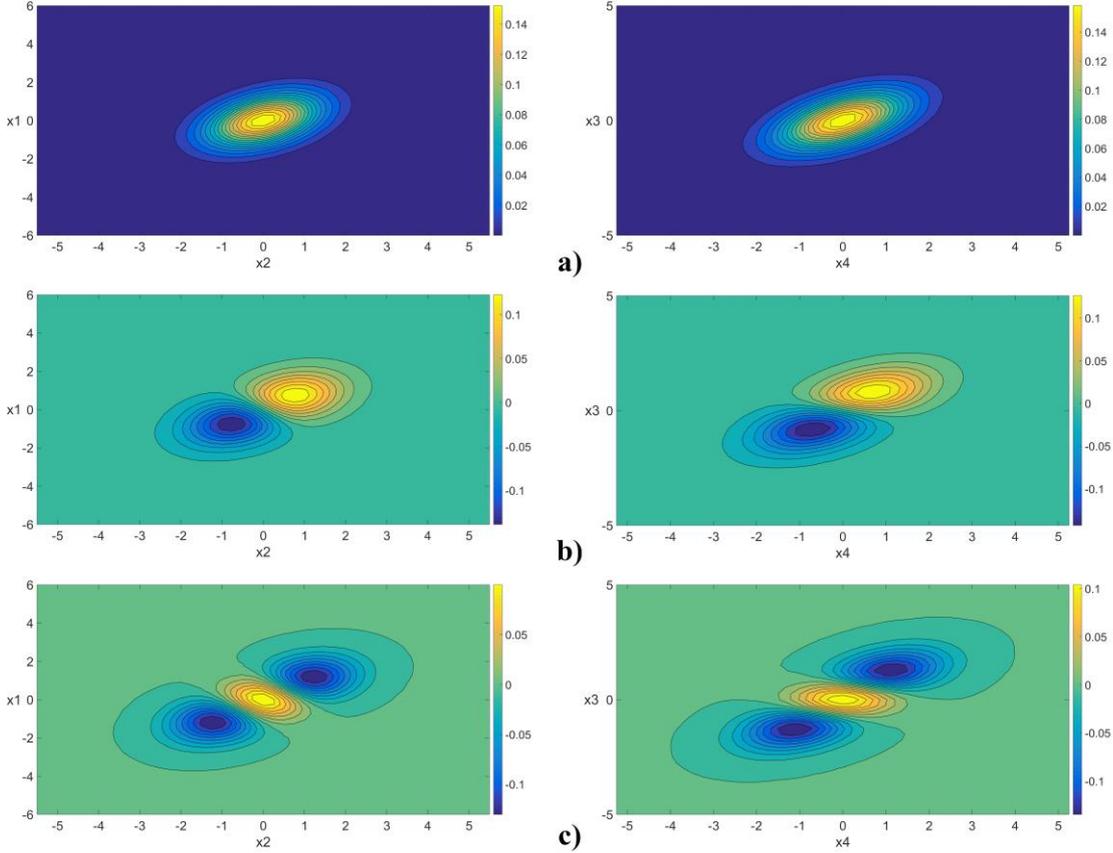

Figure 6. The first three Schmidt modes of a four-dimensional Gaussian state. To the left, Schmidt modes corresponding to the subsystem of random variables $X_1, X_2$, and to the right, Schmidt modes corresponding to the subsystem of random variables $X_3, X_4$.

Then the states of systems $A$ and $B$ are given by the following products $\prod_{j=1}^{r} \psi_{n_j}^{A}(z)$ and $\prod_{j=1}^{r} \psi_{n_j}^{B}(w)$ respectively with the same set of numbers $n_1, n_2, ..., n_r$. These states are the Schmidt modes. The weights of such mode are given by $\lambda_{n_1 n_2 ... n_r} = \prod_{j=1}^{r} \lambda_{n_j}$. Here, $\lambda_{n_j}$ is the population of the $n_j$ state, where $j$ - is the number of canonical oscillator. As noted earlier, the coefficients $\lambda_{n_j}$ (with fixed number $j$) form a geometrical progression. The obtained Schmidt modes should be sorted (starting with the modes with higher weights). In addition, it is necessary to take into consideration the constant factor which determines the Jacobian of transform to the new variables.

**4.2 Four-dimensional normal distribution**

We also considered a four-dimensional normal distribution corresponding to a system of four centered random variables $X_1, X_2, X_3, X_4$. For example, the following correlation matrix can be chosen

$$\rho = \begin{pmatrix} 1 & 0.7 & 0.8 & 0.85 \\ 0.7 & 1 & 0.9 & 0.65 \\ 0.8 & 0.9 & 1 & 0.75 \\ 0.85 & 0.65 & 0.75 & 1 \end{pmatrix}.$$

The initial probability distribution corresponds to quantum state (5) which is a four-dimensional numerical array. Having rearranged elements in this array, it is possible to get a matrix, the rows of which correspond to a subsystem of random variables $X_1, X_2$, and the columns to subsystem $X_3, X_4$. Applying the Schmidt decomposition, it is possible to find Schmidt modes for both subsystems. In this case we shall have surfaces that are presented in Fig. 6. It should be noted that numerical calculations confirm the validity of the analytical equations given in section 4.1.

The Schmidt number corresponding to this distribution is approximately equal to $K = 3.5306$. Partial Schmidt numbers are equal to $K_1 = 2.9453$, $K_2 = 1.1987$. The weights of calculated modes are equal to $\lambda_0 = 0.4611$, $\lambda_1 = 0.2274$, $\lambda_2 = 0.1121$ respectively. The Schmidt coefficient (18) coincides with Pearson multiple correlation coefficient (19) $\rho_{Pearson}^2 = \rho_{Schmidt}^2 = 0.9198$.

The considered example shows applicability of Schmidt decomposition in the correlation analysis for high dimensional systems.

## 5. CONCLUSIONS

It is shown that the Schmidt decomposition formalism is an adequate tool for the correlation dependence analysis, both classical and quantum. The analytical theory of Schmidt decomposition for a multivariate Gaussian state is developed. It is shown that mathematical tools of quantum mechanics can significantly improve the classical statistical analysis. This method makes it possible to deduce far-reaching generalizations regarding not only standard 2D Pearson correlations but also the partial and multiple multidimensional correlations. The developed formalism which is based on Schmidt decomposition can be applied to correlations problems analysis.

The research results are significant for a deeper understanding of the statistical analysis and can be used for developing high-dimensional experimental quantum systems for data processing.

The work was supported by the Russian Foundation for Basic Research (project no. 13-07-00711), and by the Program of the Russian Academy of Sciences in fundamental research.